\documentclass[twocolumn,traditabstract]{aa}
\usepackage{amssymb}
\usepackage{mathptmx}
\usepackage{natbib}
\usepackage[]{graphicx}
\defcitealias{nesvadba16}{N16}

\begin{document}
\title{Gas kinematics in powerful radio galaxies at z$\sim$2: Energy supply from star formation, AGN, and radio jet
\thanks{Based on observations carried out with the Very Large
Telescope of ESO under Program IDs 079.A-0617, 084.A-0324, 
085.A-0897, and 090.A-0614.\newline
Herschel is an ESA space observatory with science instruments provided
by European-led Principal Investigator consortia and with important
participation from NASA.}}  \author{N.~P.~H.~Nesvadba\inst{1,2,3}
G.~Drouart\inst{4}, C. De Breuck\inst{5}, P.~Best\inst{6}, N.~Seymour\inst{4}, J.~Vernet\inst{5}}
\institute{Institut d'Astrophysique
Spatiale, UMR8617, Universit\'e Paris-Sud, b\^at 121, Orsay, France
d'Orsay, Bat.~121, 91405 Orsay, France
\and
CNRS, Orsay, France
\and 
email: nicole.nesvadba@ias.u-psud.fr
\and
International Center for Radio Astronomy Research, Curtin University, Perth, Australia
\and
European Southern Observatory, Karl-Schwarzschild Strasse 2, 85748 Garching bei M\"unchen, Germany
\and 
SUPA, Institute for Astronomy, Royal Observatory of Edinburgh, Blackford Hill, Edinburgh, EH9 3HJ, UK
}
\titlerunning{Gas kinematics in HzRGs}
\authorrunning{Nesvadba et al.}  \date{Received / Accepted }

\date{Accepted . Received ; in original form }

\abstract{
We compare the kinetic energy and momentum injection rates from
intense star formation, bolometric AGN radiation, and radio jets with
the kinetic energy and momentum observed in the warm ionized gas in 24
powerful radio galaxies at z$\sim$2. These galaxies are amongst our
best candidates for being massive galaxies near the end of their
active formation period, when intense star formation, quasar activity,
and powerful radio jets all co-exist. All galaxies have VLT/SINFONI
imaging spectroscopy of the rest-frame optical line emission, showing
extended emission-line regions with large velocity offsets (up to
1500~km s$^{-1}$) and line widths (typically 800-1000 km s$^{-1}$)
consistent with very turbulent, often outflowing gas. As part of the
HeRG\'E sample, they also have FIR estimates of the star formation and
quasar activity obtained with Herschel/PACS and SPIRE, which enables
us to measure the relative energy and momentum release from each of
the three main sources of feedback in massive, star-forming AGN host
galaxies during their most rapid formation phase.  We find that star
formation falls short by factors 10$-$1000 of providing the energy and
momentum necessary to power the observed gas kinematics. The obscured
quasars in the nuclei of these galaxies provide enough energy and
momentum in about half of the sample, however, only if these are
transfered to the gas relatively efficiently. We compare with
theoretical and observational constraints on the efficiency of the
energy and momentum transfer from jet and AGN radiation, which
advocates that the radio jet is the main driver of the gas kinematics.}


\maketitle

\section{Introduction}
\label{sec:introduction}

Powerful radio galaxies (HzRGs) at high redshift (z$\ga$2) are ideal
targets to study the late formation stages of massive galaxies in the
early Universe. They have high stellar \citep[e.g.,][]{debreuck01,
  seymour07, debreuck10} and dynamical masses \citep[][]{villar03,
  nesvadba07b}, and often high star-formation rates of up to 1000
M$_{\odot}$ yr$^{-1}$ \citep[][]{archibald01, reuland04, drouart14},
with implied formation times of few 100~Myr. They host luminous,
obscured quasars \citep[e.g.,][]{carilli02, overzier05, drouart14},
and have powerful radio jets \citep[e.g.,][]{carilli97,pentericci00},
indicating that they are the host galaxies of some of the most
powerful active galactic nuclei. Their black hole masses fall near the
upper end of the mass function of supermassive black holes in nearby
galaxies \citep[][]{nesvadba11a}, and scale with the mass of their
host galaxies in a fairly similar way to nearby galaxies which fall
onto the local black-hole bulge mass relationship, suggesting they
must be near the end of their active formation
period. \citet{drouart14} argued that the black holes of HzRGs will
outgrow the plausible mass range for supermassive black holes even in
very massive galaxies, if their growth will continue for more than few
$10^7$ yrs, further highlighting that we are observing these
sources at a critical moment of their evolution. The same is
suggested by their high stellar masses, which exceed the amounts of
remaining molecular gas by factors 10 or more \citep[][]{seymour07,
  debreuck10, emonts14}, limiting their potential future growth in
stellar mass.

HzRGs are often surrounded by extended nebulosities of high
surface-brightness, warm ionized gas \citep[e.g.,][]{villar03,
  nesvadba08} with sizes of up to about 60~kpc, and irregular gas
kinematics, with velocity offsets and line FWHMs of up to 1000~km
s$^{-1}$. In very powerful HzRGs, these velocities are above the
escape velocity from the gravitational potential of massive galaxies,
suggesting this gas is outflowing \citep[][]{villar03, nesvadba06a,
  nesvadba08}. In galaxies with more moderate jet power, smaller-scale
outflows and turbulence seem to be dominant \citep[][N16
  hereafter]{collet15b,nesvadba16}. The high surface-brightness
emission line regions are in most cases elongated along the radio jet
axis, and their sizes typically do not exceed the jet size.  Dynamical
times are comparable to the radio-jet lifetime. This supports the
interpretation that these are outflows of ambient, warm ionized gas,
which has been entrained by the expanding 'cocoon' of hot, shocked gas
inflated by the radio jet \citep[][]{nesvadba06a, nesvadba08}.

However, powerful radio galaxies are complex environments, where star
formation and bright quasar activity co-exist with the powerful radio
jets, and it has so far been impossible to compare the possible
relative contribution from star formation and quasar activity to the
gas kinematics in these systems directly. Star formation as a possible
driver of extended, super-galactic bubbles has been suggested by,
e.g., \citet{mccarthy99}, \citet{taniguchi01}, \citet{zirm05}, and
\citet{humphrey09}. Some of this star formation could be extended in
the halo surrounding the radio galaxy itself \citep[e.g.,][]{hatch08,
  hatch13}.

Estimating star-formation rates in high-redshift radio galaxies is
observationally challenging. Common star formation tracers like the UV
continuum, bright optical emission lines like H$\alpha$, or rest-frame
mid-infrared photometry probing PAHs are generally not reliable
tracers of star formation in AGN host galaxies, because they are
either contaminated or even dominated by heating through AGN photons
and shocks. Although 850~$\mu$m photometry with SCUBA has been taken in
the past, and provided interesting qualitative evidence of star
formation in HzRGs, deriving robust quantitative constraints has been
difficult because of the uncertain spectral shapes and dust
temperatures in HzRGs. In addition, the large beam of single-dish
measurements raises the dangers of confusing star formation within the
radio galaxy with star formation in nearby companion galaxies that
fall within the same beam \citep[e.g.][]{debreuck01, ivison08,
  nesvadba09a, drouart14, emonts14}. This risk is particularly high in
HzRGs since these are predominantly probing dense environments,
presumably forming galaxy clusters \citep[][]{hatch09}. Furthermore, by
analogy with Spitzer and Herschel observations of nearby radio
galaxies, we may suspect that even at rest-frame wavelengths of
$100-200~\mu$m, as probed in the 850~$\mu$m band in galaxies at
redshifts $z\sim 2-3$, the FIR continuum can still be dominated by the
AGN, not the star formation \citep[][]{tadhunter14}.

Having a sample of HzRGs with good mid to far-infrared coverage to
disentangle dust heating through AGN radiation and star formation, is
therefore necessary to study the role of feedback from each process in
powerful high-redshift radio galaxies. In addition, we need good
constraints of the gas kinematics, ideally over the whole projected
surface of the nebulosities.

\begin{figure}
\begin{center}
\includegraphics[width=0.49\textwidth]{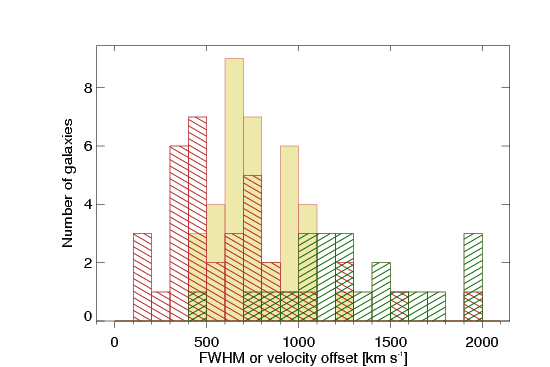}
\caption{Histograms of the average (yellow
  shaded histogram) and maximal (dark green hatched histogram) full
  widths at half maximum of our sources, and of their velocity offsets
  (dark red hatched histogram).\label{fig:histokine}}
\end{center}
\end{figure}

We present an analysis of the gas energetics in a sample of 24~HzRGs
at z$\ga$2, which have good infrared constraints of obscured AGN and
star-formation activity, and also rest-frame optical imaging
spectroscopy of warm ionized gas. These galaxies are the subset of
sources from the HeRG\'E project \citep[][]{drouart14,seymour12},
which are also part of our SINFONI study of 49 HzRGs at z$\ga$2
\citep[][N16]{nesvadba06a, nesvadba07b, nesvadba08b, nesvadba11a,
  collet15a, collet15b}. Our sources have AGN luminosities in the
range of $L_{\rm bol,AGN}=$few$\times 10^{13}$ $L_{\rm \odot}$ and
radio power of $\ge 10^{27}$ W Hz$^{-1}$. They are amongst the most
luminous obscured, radio-loud quasars, and are also amongst the best
studied radio galaxies at their epoch. In our previous work we
provided a number of observations suggesting that the jet most likely
dominates the kinetic energy injection into the gas, including a close
alignment between the jet axis and the kinematic and morphological
major axis of the gas, and velocity jumps or elevated line widths near
bright radio features in some galaxies. However, as encouraging as
these findings are, it is not sufficient to show that the jet may
plausibly accelerate the gas, we also need to show that alternative
mechanisms are less likely. This has now become possible thanks to the
HeRG\'E constraints, and is the goal of the present study. 

\begin{figure}
\centering
\includegraphics[width=0.45\textwidth]{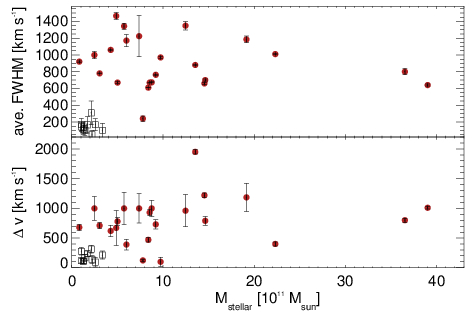}
\caption{Emission-line FWHM {\it (top)} and
  velocity offset {\it (bottom)} as a function of stellar mass, as
  estimated by \citet{seymour07} and \citet{debreuck10} from Spitzer
  near-to-mid-infrared photometry. Red dots show the galaxies from our
  sample, the small black empty squares the stellar-mass selected
  sample of \citet{buitrago14} without powerful AGN for comparison. We
  do not find a correlation for our sample in either
  relationship. \label{fig:kinevsmstellar}}
\end{figure}

\citet{seymour12} have already carried out a similar analysis for one
galaxy of our set. For the ``Spiderweb'' galaxy MRC~1138$-$262 at
z$=$2.16, they estimated a bolometric power of the AGN of $L_{\rm
 bol,AGN}\sim 7\times 10^{13}\ L_{\odot}$ and argued that this would
suffice to power the outflow with kinetic energy of $L_{\rm kin}\sim
8\times 10^{12} L_{\rm \odot}$ observed in MRC~1138$-$262 by
\citet{nesvadba06a}. Hence the AGN as well as the radio jet would seem
energetically capable to drive the gas kinematics in this galaxy, if
of-order 10\% of the bolometric luminosity of the AGN were transformed
into kinetic energy of the gas. In turn, the kinetic energy released
by the star formation in MRC~1138$-$262, although one of the most
vigorously star-forming high-redshift radio galaxies known, would not
be sufficient \citep[][]{nesvadba06a, seymour12}.

\begin{figure*}
\begin{center}
\includegraphics[width=0.43\textwidth]{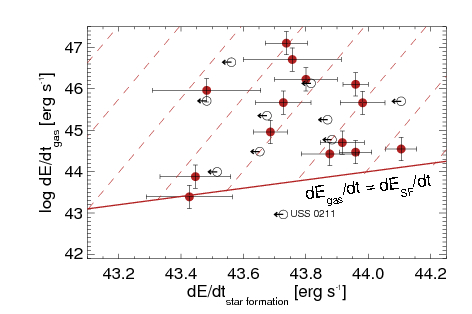}
\includegraphics[width=0.43\textwidth]{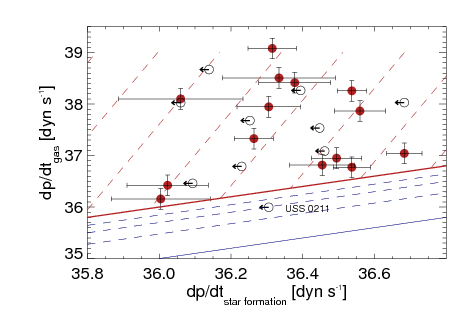}
\includegraphics[width=0.43\textwidth]{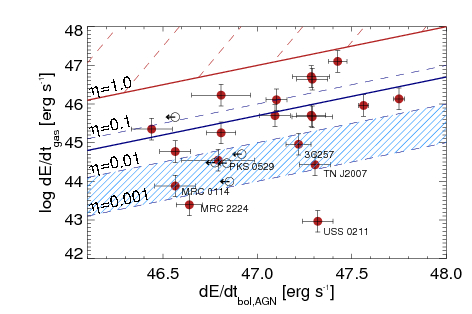}
\includegraphics[width=0.43\textwidth]{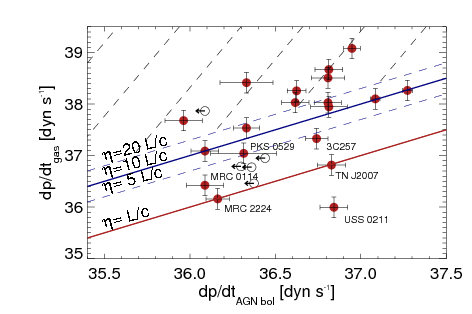}
\includegraphics[width=0.43\textwidth]{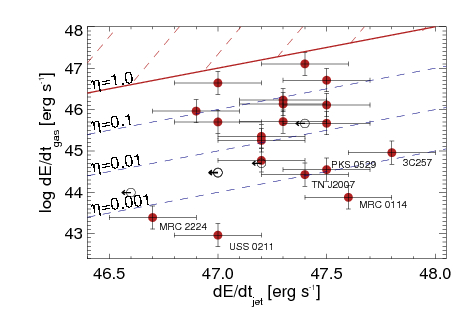}
\includegraphics[width=0.43\textwidth]{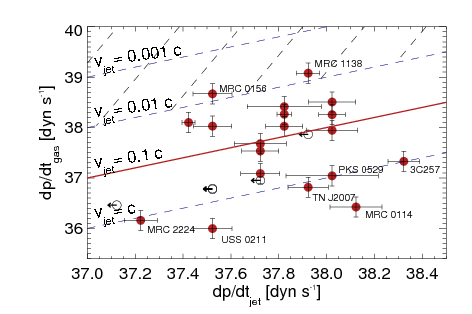}
\caption{Relationships between the observed kinetic energy (left)
    and momentum (right) of the gas, for star formation (top), AGN
    radiation (center) and radio jets (bottom). Filled red and empty
    black dots show galaxies with FIR detections and upper limits,
    respectively. Error bars show the measurement uncertainties. Red
    wide-spaced dashed regions are excluded by energy and momentum
    conservation, black wide-spaced dashed regions by empirical
    constraints on the maximal momentum transfer from AGN radiation
    and radio jet, respectively. In the top right panel, the red and
    blue solid lines show momentum transfer of 100 and 10\% of that
    provided by star formation, respectively. \citet{heckman15} argued
    that 10\% efficiency is most likely. Thin dashed dark blue lines
    show transfer rates of 30, 50, and 70\%. In the middle left panel,
    the red line shows an energy transfer of 100\%, the blue line of
    5\% required by cosmological models, the hatched region of 0.1-1\%
    of $L_{bol}$ to gas kinetic energy. The latter are favored by
    observations of FeLoBALs \citep[][]{moe09, dunn10, bautista10} and
    hydrodynamic models taking radiative transfer explicitly into
    accound \citep[][]{bieri16, krumholz12}. The same observations and
    models also favor an average momentum transfer of $L/c$ (red line
    in the central right panel), whereas more optimistic models find
    $L/c\sim 10$ \citep[dark blue solid line;][]{zubovas12, faucher12}. In
    the bottom left panel, the red solid line represents an energy
    transfer of 100\% from jet to gas, and in the bottom right panel the
    momentum transfer for the most likely jet expansion velocity of
    0.1c. Dashed dark-blue lines show the momentum input from the jet
    for different assumptions of the jet advance speed, assuming that
    no momentum is lost to other effects. It is therefore not
    impossible to find galaxies below the line for a jet expansion at
    the speed of light, $c$ (lowest dark-blue dashed line), due to,
    e.g., geometry, gas clumpiness, and other effects.
 \label{fig:sfratio}}
\end{center}
\end{figure*}

Building upon these results for a first galaxy, we present a direct
comparison of the energy injection rates from star formation, quasar
bolometric power and radio jet. We expand the previous results by
including all 24~galaxies with Herschel photometry as well as
VLT/SINFONI imaging spectroscopy, and by analyzing the energy as well
as the momentum injection into the gas. We discuss our results in
light of recent feedback models.  

Throughout the paper we adopt a flat ${\rm H_0}= 70\,{\rm km}\,{\rm
 s}^{-1}\,{\rm Mpc}^{-1}$ concordance cosmology with $\Omega_{\rm
 M}=0.3$ and $\Omega_{\Lambda}=0.7$.

\begin{table*}
\centering
\begin{tabular}{lccccccc}
\hline
Source          & RA & Dec & z    & dE/dt$_{\rm gas}$  & dE/dt$_{\rm AGN,bol}$ & dE/dt$_{\rm SF}$ & dE/dt$_{\rm jet}$ \\
                & (J2000) & (J2000) &  & [erg s$^{\rm -1}$] & [erg s$^{\rm -1}$] & [erg s$^{\rm -1}$] & [erg s$^{\rm -1}$] \\
\hline
MRC~0114$-$211  & 01:16:51.4 & $-$20:52:07 & 1.41 & $43.9\pm$0.3 & $46.6\pm$0.1 & $43.4\pm$0.1    & $47.6\pm0.3$ \\
TNJ~0121$+$1320 & 01:21:42.7 & $+$13:20:58 & 3.52 & $44.5\pm$0.3 & $<46.9$     & $44.0\pm$0.1    & $47.0\pm0.3$ \\
MRC~0156$-$252  & 01:58:33.6 & $-$24:59:31 & 2.02 & $46.6\pm$0.3 & $47.3\pm$0.1 & $<43.6$ & $47.0\pm0.3$ \\
TNJ~0205$-$2242 & 02:05:10.7 & $+$22:42:50 & 3.51 & $44.5\pm$0.3 & $<46.8$ & $<43.7$ & $47.0\pm0.3$ \\
MRC~0211$-$122  & 02:14:17.4 & $-$11:58:46 & 2.34 & $43.0\pm$0.3 & $47.3\pm$0.1 & $43.7\pm$0.2    & $47.0\pm0.3$ \\
MG~0251$-$273   & 02:53:16.7 & $-$27:09:13 & 3.16 & $45.2\pm$0.3 & $46.8\pm$0.1 & $<43.9$ & $47.2\pm0.3$ \\
MRC~0316$-$257  & 03:18:12.1 & $-$25:35:10 & 3.14 & $45.7\pm$0.3 & $<46.6$ & $44.0\pm$0.1     & $47.4\pm0.3$ \\
MRC~0406$-$244  & 04:08:51.4 & $-$24:18:17 & 2.44 & $46.1\pm$0.3 & $47.1\pm$0.1 & $44.0\pm$0.1    & $47.5\pm0.3$ \\
PKS~0529$-$549  & 05:30:25.4 & $-$54:54:22 & 2.58 & $44.5\pm$0.3 & $46.8\pm$0.1 & $44.1\pm$0.1    & $47.5\pm0.3$ \\
5C~7.269        & 08:28:38.8 & $+$25:28:27 & 2.22 & $44.0\pm$0.3 & $<46.9$ & $<43.5$ & $46.6\pm0.3$ \\
TXS~0828$+$193  & 08:30:53.4 & $+$19:13:16 & 2.57 & $46.0\pm$0.3 & $47.6\pm$0.1 & $43.5\pm$0.2    & $46.9\pm0.3$ \\
3C257           & 11:23:09.4 & $+$05:30:18 & 2.46 & $45.0\pm$0.3 & $47.2\pm$0.1 & $43.7\pm$0.1    & $47.8\pm0.3$ \\
MRC~1138$-$262  & 11:40:48.3 & $-$26:29:10 & 2.16 & $46.5\pm$0.3 & $47.4\pm$0.1 & $43.7\pm$0.1    & $47.4\pm0.3$ \\
USS~1243$+$036  & 12:45:38.4 & $+$03:23:21 & 3.57 & $46.7\pm$0.3 & $47.3\pm$0.1 & $43.8\pm$0.2    & $47.5\pm0.3$ \\
USS~1410$-$001  & 14:13:15.1 & $-$00:23:00 & 2.36 & $45.7\pm$0.3 & $47.1\pm$0.1 & $<44.1$0.1 & $47.0\pm0.3$ \\
MRC~1558$-$003  & 16:01:17.3 & $-$00:28:48 & 2.53 & $45.7\pm$0.3 & $47.3\pm$0.1 & $<43.5$ & $47.3\pm0.3$ \\
USS~1707$+$105  & 17:01:06.5 & $+$10:31:06 & 2.35 & $45.4\pm$0.3 & $46.5\pm$0.1 & $<43.7$ & $47.2\pm0.3$ \\
TNJ~2007$-$1316 & 20:07:53.2 & $-$13:16:44 & 3.84 & $44.2\pm$0.3 & $47.3\pm$0.1 & $43.9\pm$0.1    & $47.4\pm0.3$ \\
MRC~2025$-$218  & 20:27:59.5 & $-$21:40:57 & 2.63 & $44.8\pm$0.3 & $<46.6$ & $<43.9$ & $47.2\pm0.3$ \\
MRC2104$-$242   & 21:06:58.2 & $-$24:05:11 & 2.49 & $46.2\pm$0.3 & $46.8\pm$0.2 & $43.8\pm$0.1   & $47.3\pm0.3$ \\
4C23.56         & 21:07:14.8 & $+$23:31:45 & 2.48 & $46.1\pm$0.3 & $47.8\pm$0.1 & $<43.8$ & $47.3\pm0.3$ \\
MRC~2144$+$1928 & 21:44:07.5 & $+$19:29:15 & 3.59 & $45.7\pm$0.3 & $47.3\pm$0.1 & $43.7\pm$0.1    & $47.5\pm0.3$ \\
MRC2224$-$273   & 22:27:43.2 & $-$27:05:02 & 2.15 & $43.4\pm$0.3 & $46.7\pm$0.1 & $43.4\pm$0.1   & $46.7\pm0.3$ \\
\hline
\end{tabular}
\caption{Coordinates, redshifts, gas kinetic energy, AGN bolometric
  power, and kinetic energy from star formation and radio
  jet for the sources in this sample \label{tab:energy}. FIR luminosities from AGN and star formation
  have previously been given by \citet{drouart14}, and 1.4~GHz radio power in
  \citetalias{nesvadba16}, respectively.}
\end{table*}

\section{Mid-to far-infrared SEDs and IR luminosities}
\label{sec:sample}

The set of galaxies we study here is a subset of the HeRG\'E set of 70
distant radio galaxies at redshifts $1<{\rm z}<5.2$ with Herschel PACS
and SPIRE photometry in the far-infrared
\citep[][]{drouart14,seymour12}. They were also observed with SINFONI,
and are therefore at redshifts where bright emission lines fall into
the near-infrared atmospheric windows \citepalias[predominantly
  z$\sim$2.0-2.6, and z$\sim3.0-3.6$][]{nesvadba16}, and at either
southern or equatiorial declinations. All galaxies also have Spitzer
mid-infrared photometry \citep[][]{seymour07,debreuck10}, and sub-mm
photometry obtained with either JCMT/SCUBA
\citep[][]{archibald01,reuland04} at 850~$\mu$m, or APEX/LABOCA
\citep[][]{drouart14} at 870~$\mu$m. In total, their
infrared-to-sub-millimeter spectral energy distributions cover the
wavelength range between 16 and 870~$\mu$m. 

\citet{drouart14} used
{\tt DecompIR} \citep[][]{mullaney11} to decompose these SEDs into an
AGN and a starburst component, using the empirical templates of
\citet{mullaney11} for both components. \citet{mullaney11} provide
five different empirical templates for the starburst component. Most
of the galaxies in our present sample are best fit by their SB~2
template, i.e., the infrared spectral energy distribution of NGC~7252
\citep[][their Table~6]{drouart14}. The 24 sources we are concerned
with here have infrared luminosities from AGN and star formation
between 1.6 and $24.5\times 10^{12}\ {\rm L}_{\odot}$ and 2.2 and
$10.5\times 10^{12}\ {\rm L}_{\odot}$, respectively \citep[Table~6
  of][]{drouart14}, integrating over wavelengths $8-1000\,\mu$m in the
rest-frame. Spectral decompositions for individual sources are
  shown in Fig.~D.1 of \citet{drouart14}.

\section{Kinetic energy and momentum of the warm ionized gas}
\label{sec:gaskinematics}
All galaxies have also been observed with the SINFONI imaging
spectrograph on the Very Large Telescope of ESO
\citep[][N16]{nesvadba06a,nesvadba08,collet15b}. SINFONI is an image
slicer with 8\arcsec$\times$8\arcsec\ field of view operating in the
near-infrared J, H, and K bands with spectral resolving power between
R=1500 and 3000. We mainly used the seeing limited mode with spatial
sampling of 125~mas$\times$250~mas to observe the bright rest-frame
optical emission lines, in particular [OIII]$\lambda\lambda$4959,5007
and H$\alpha$, depending on redshift.
\begin{figure*}
\begin{center}
\includegraphics[width=0.49\textwidth]{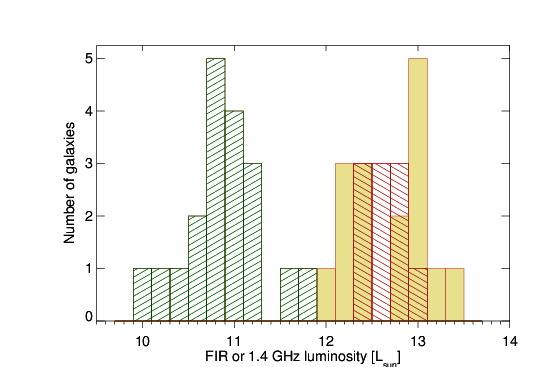}
\includegraphics[width=0.49\textwidth]{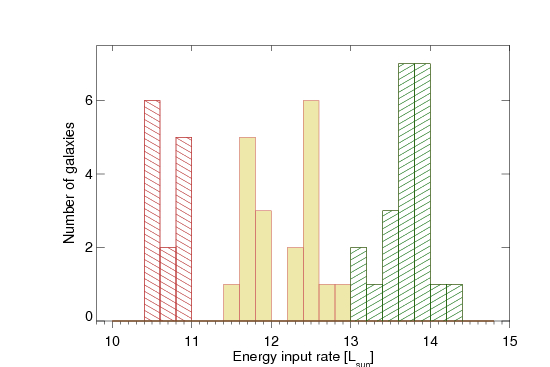}
\caption{\label{fig:histoluminosities}{\it left} 
  Histograms of the FIR luminosity between 8 and 1000 $\mu$m, and the
  1.4~GHz monochromatic power, i.e., the quantities used to estimate
  the kinetic energy of the different possible drivers discussed in
  this paper. The yellow filled, red hatched and green hatched
  diagrams correspond to bolometric AGN power, star formation, and
  radio power, respectively. We only include sources with Herschel
  detections.{\it right} The same histograms showing the kinetic
  energy corresponding to the FIR luminosity and monochromatic radio
  power at 1.4~GHz, derived with the methods described in
  \S\ref{ssec:starformation} and \S\ref{ssec:lbol}.}
\end{center}
\end{figure*}

Fig.~\ref{fig:histokine} summarizes the kinematic properties of the
warm ionized gas in our galaxies, which has already been presented in
more detail by \citetalias{nesvadba16}, in the form of three histograms:
we show the average and maximal full-width-at-half-maximum (FWHM) measured
from [OIII]$\lambda$5007 in our SINFONI cubes for each galaxy,
respectively, and the maximal velocity offset in the extended
emission line gas. The average FWHM ranges from FWHM$_{\rm
  avg}=400$~km s$^{-1}$ to 1100 km s$^{-1}$, and reaches 2500 km
s$^{-1}$ in USS~1243$+$036. It represents the average of all local
FWHMs measured typically from [OIII]$\lambda$5007 in all spatial
pixels in a given galaxy where the signal-to-noise ratio exceeds
three. The maximal FWHM measured in small regions of our sources are
between 700~km~s$^{-1}$ and 1950~km~s$^{-1}$, and are typically about
500 km s$^{-1}$ greater than the average FWHM
\citepalias[][]{nesvadba16}. The velocity offsets are between 100~km
s$^{-1}$ and 1300 km s$^{-1}$ for most sources. We note that even if
we corrected the velocity offsets by fiducial factors 2$-$3 to account
for inclination effects, the radial velocities of the gas would
typically not exceed the FWHM line widths.

A potential caveat of these estimates might be that
[OIII]$\lambda$5007 is probing a highly ionized state of a trace
element, and may therefore not be a representative tracer of the
dominant mass component. However, \citetalias{nesvadba16} compared the
velocity fields of [OIII]$\lambda$5007 and H$\alpha$ in the nine
galaxies where both are emitted from spatially well extended regions
and at good signal-to-noise ratios, finding little cause for alarm, at
least at the spatial resolutions of about 5-10~kpc reached with this
study.

Our imaging spectroscopy data provide several constraints that we can
use to estimate the kinetic energy in the warm ionized gas of our
sources. We follow \citet{nesvadba06a} in estimating the energy
  necessary to inflate a hot bubble adiabatically to the sizes, r, and
  velocities, $\Delta v$, as observed:
\begin{equation}
dE/dt = 1.5\times 10^{46}\times r^2\times \Delta v^3\ \times n_0\ {\rm erg\ s^{-1}}
\end{equation}
where $r$ is the radius of the bubble in units of 10~kpc. $\Delta v$
is given in units of 1000~km s$^{-1}$, and we adopt $n_0$ = 1
cm$^{-3}$ for the density of the gas into which the bubble expands.

This estimate does not depend on the total gas mass, and is therefore
not affected by gas in other phases than the warm ionized phase, which
may contribute significantly to the total mass in the outflow
\citep[e.g.,][]{cicone14}. The only requirement is that the gas
kinematics are representative for the terminal outflow velocity, which
should be the case, because they subtend large volumes surrounding the
galaxy relatively evenly and have moderate densities compared to the
colder, and perhaps dominant, gas phases
\citepalias[][]{nesvadba16}. Table~\ref{tab:energy} shows
that the resulting kinetic energy in our targets is between $\log
E_{gas}=43.0$ and $47.5$ erg s$^{-1}$.

\section{Stellar masses and gravitational motion}
\label{ssec:gravitationalmotion}

Before turning to the potential astrophysical mechanisms that may
power the gas kinematics in our HzRGs, we will first present our
arguments why gravitational motion is unlikely to play a large
role. \cite{seymour07} and \citet{debreuck10} used Spitzer and
ground-based photometry to decompose the optical-to-infrared spectral
energy distributions of our sources into their stellar and AGN dust
components, obtaining robust stellar mass estimates (or at least upper
limits) for most. Most HzRGs fall into a small range of (high) stellar
masses around $\log{{\rm M_{stellar}/M_{\odot}}}=11.5\pm 0.3$. 

In Fig.~\ref{fig:kinevsmstellar}, we show the total velocity offsets
and average FWHMs of the gas as a function of the stellar mass derived
by \citet{seymour07} and \citet{debreuck10}, for the sources we study
here.  We find no evidence of a relationship between stellar mass and
velocity or FWHM, as measured across the extended emission-line
regions, and consider this additional evidence that gravity does not
play a large role for the gas motion in our sources. We also show
  the sources of \citet{buitrago14}, to our knowledge the only
  available stellar mass selected sample of galaxies in the early
  Universe with rest-frame optical imaging field spectroscopy and with
  $M_{stellar}>10^{11} M_{\odot}$. This adds to the arguments already
presented by \citet{collet15b} and \citetalias{nesvadba16}, namely:
(1) that the velocity and dispersion maps of the gas are very complex,
inconsistent with simple rotating disks; (2) line widths are
systematically and significantly greater than in quiescent high-z
bulges with similar stellar mass; (3) their ratios of
velocity gradients to line widths are inconsistent with the
ellipticities for both fast and slow rotators in the $Atlas^{3D}$
sample of nearby galaxies.

The broad line widths and high velocities further suggest that an
additional energy and momentum input mechanism must be present in the
HzRGs. In the following sections we will use our observational
constraints on star-formation rates, AGN bolometric power, and radio
jets, to further discuss which of these three dominates.

\section{The power of star formation}
\label{ssec:starformation}
Intense star formation in galaxies releases kinetic energy and
momentum into the ambient gas in form of stellar winds from young
stars, as well as through expanding shells of supernova remnants,
which interact, mix, and thermalize, producing smooth, galaxy-wide
outflows \citep[e.g.,][]{heckman90, lehnert96a,
  lehnert96b}. Observations and simulations have now reached a broad
consensus that about 40\% of the initial $10^{51}$ erg of kinetic
energy released by a supernova is carried by galactic winds
\citep[e.g.,][]{dallavecchia08, sharma14, veilleux05, strickland09},
thus releasing about $1-2\times 10^{49}$ erg of kinetic energy per solar
mass of stars formed into the interstellar medium
\citep[][]{dallavecchia08}. The precise estimate depends most strongly
on the assumptions about the initial mass function of star formation
and details of the thermalization and mass loading process. Here
 we adopt an energy release of $10^{49}$ erg per solar mass of star
 formation.

We estimate the kinetic energy released by star formation in our HzRGs
from the starburst component of the far-infrared luminosity measured
by \citet{drouart14}, and calculate a star-formation rate by setting
${\rm SFR =} 4.5\times 10^{-44}\ L_{FIR}$ \citep[][]{kennicutt98}.
The SFR is given in M$_{\odot}$ yr$^{-1}$, and the FIR luminosity in
erg s$^{-1}$. \citet{kennicutt98} derived this relationship for a
  Salpeter initial mass function; we divide the star-formation rates
  by a factor 1.8 to match the Chabrier initial mass function more
  commonly used today \citep[][]{chabrier03}.

With star-formation rates between 211 and 550~M$_{\odot}$ yr$^{-1}$,
and using the above mentioned scaling that supernovae release on
average $1\times 10^{49}$ erg of kinetic energy per solar mass of
stellar mass formed, we find kinetic energies between $1.7\times
10^{44}$ erg s$^{-1}$ and $2.2\times 10^{45}$ erg s$^{-1}$. This
corresponds to $0.4-5.8\times 10^{11}$ L$_{\odot}$. 

To determine the momentum produced by the star formation in our
sources, we rely on the recent empirical study of powerful
starburst-driven winds in low-redshift galaxies by \citet{heckman15},
who find that the momentum injection per solar mass of new stars
formed is about $4.8\times 10^{33}$ dyn. This includes the
contribution of ram pressure from the hot wind medium as well as that
of radiation pressure. They also find that the starburst must inject
about 10$\times$ more momentum into the gas than is necessary to
unbind the wind, in order to produce clear outflow signatures. In our
targets this corresponds to $9\times 10^{35}$ dyn to $6\times 10^{36}$
dyn. Values for individual galaxies are listed in
Table~\ref{tab:energy}.

Fig.~\ref{fig:sfratio} shows the kinetic energy and momentum of the
gas as a function of the energy and momentum that are released by star
formation with the above estimates. Nearly all sources are above the
solid black line for at least one quantity, with only one exception,
USS~$0211-122$ at $z=2.1$. This galaxy has unusually narrow
emission lines for our overall sample, and a distinctive unresolved
broad component, which is also unusual for our sample
\citepalias[][]{nesvadba16}. \citet{vernet01} also found it was an
outlier compared to other HzRGs, with unusually high levels of
polarization, consistent with low levels of star formation, and high
nitrogen abundance, suggesting high levels of secondary nitrogen
production, and perhaps a very advanced evolutionary state.

The remaining 23 galaxies require more energy and momentum input than
can be provided by star formation, by up to three orders of
magnitude, even if this energy and momentum were transferred into the
interstellar medium at 100\% efficiency. This is likely not a
realistic assumption. Observations of starburst-driven winds suggest
that efficiencies are more in the range of a few 10\%
\citep[][]{strickland09} for the energy, and about 10\% for the
momentum transfer \citep[][]{heckman15}, larger than the
  potential systematic uncertainties. The latter efficiency is shown
as dashed black line in the top right panel of
Fig.~\ref{fig:sfratio}. Many theoretical studies suggest values that
are even lower \citep[e.g.,][]{krumholz12, bieri16}. As an additional
caveat, recent ALMA arcsec-resolution observations have shown that
most of the FIR emission from star formation does not necessarily come
from the radio galaxy itself, but from nearby companion galaxies
\citep[][see also \citealt{ivison08}, N16]{gullberg16}. This would
also suggest that we might overestimate the energy and momentum input
rates shown in Fig.~\ref{fig:sfratio}. Taking these arguments into
account only reinforces our finding that star formation in these
galaxies does not produce sufficient kinetic energy or momentum to
power the gas kinematics as observed.

\section{Bolometric AGN radiation}
\label{ssec:lbol}

Observations of broad quasar absorption lines have provided secure
evidence for fast outflows within the nuclei of powerful AGN for
several decades, but observational evidence, that such outflows may
also extend into the kpc range away from galactic nuclei is being
found only now \citep[e.g.,][]{liu13a, lui13b, sun14}, although
  observations of broad, blueshifted emission-line wings from the
  narrow line region have been known for much longer
  \citep[e.g.,][]{heckman84,greene05,woo16}.  The bolometric
luminosity emitted by powerful quasars during their lifetime
corresponds approximately to the binding energy of their massive host
galaxy \citep[e.g.,][]{silk98}. Quasars might therefore in principle
drive large-scale outflows, if the radiated energy and momentum are
deposited in the ambient gas at high enough efficiency.  Cosmological
models of galaxy evolution suggest that about 0.2$-$0.5\% of the
rest-mass energy equivalent of a supermassive black hole must be
injected to reproduce the observed black-hole bulge (mass)
relationships \citep[e.g.,][]{dimatteo05}, corresponding to about
2-5\% of the bolometric luminosity of the AGN for a radiative
accretion efficiency onto the black hole of 0.1.

To estimate the bolometric quasar power, we rely on the AGN component
of the FIR spectral energy distributions of \citet[][their
  Table~6]{drouart14}, and multiply by a factor~6 \citep[again,
  following][]{drouart14}, to find the bolometric quasar luminosities
listed in Table~\ref{tab:energy} of between $6.6\times 10^{45}$ and
$5.6\times 10^{47}$ erg s$^{-1}$. These estimates are 
uncertain by factors of a few \citep[][]{drouart14}, for two main
reasons: Firstly, the AGN torus model adopted by \citet{mullaney11}
assumes relatively high dust temperatures, which would maximize the
FIR AGN luminosity by a factor $\sim 2$, at the expense of the FIR
luminosity from star formation, which would be underestimated by
about a similar factor. However, if the torus component is not
sufficiently extended, or if a significant part of the star formation
occurs in companion galaxies, we would overestimate the starburst, and
underestimate the AGN component \citep[][]{drouart16}. Secondly, as
already pointed out by \citet[][their appendix~C]{drouart14}, the
adopted AGN spectral shape used for the bolometric correction is also
uncertain by factors of a few. We adopted their favored bolometric
correction factor~6 obtained from the classical QSO template by
\citet{elvis94}; the template of \citet{richards06} would have
resulted in a bolometric correction factor of~5. For four galaxies
from the present sample \citet{nesvadba11a} estimated bolometric
luminosities from the luminosity and width of the broad nuclear
H$\alpha$ line, finding values between $\log L_{bol} = 46.6$ and
$46.8$ erg s$^{-1}$, consistent with X-ray constraints in the two
cases where X-ray constraints were available. Generally, these
estimates are in the lower range of the $\log L_{bol}=46.6$ to 47.4
erg~s$^{-1}$ found with the approach of \citet{drouart14}. Therefore,
we cannot exclude that we overestimate our bolometric AGN luminosities
by factors of a few. In the present case, however, we wish to
distinguish between the impact of quasar radiation and radio jets, so
that this is a conservative choice, which maximizes the potential
impact of the quasar radiation.

In Fig.~\ref{fig:histoluminosities} we compare the far-infrared
luminosities from the quasar and from star formation in our targets,
finding that both are quite similar, about few $10^{12}$ L$_{\odot}$
for most targets. It is however also interesting to compare the
kinetic power released by the starburst, estimated in the way
described in \S\ref{ssec:starformation} with that from the AGN,
assuming that 5\% of the bolometric AGN power is turned into
mechanical energy of the gas. Fig.~\ref{fig:histoluminosities} shows
that the kinetic energy from the AGN radiation exceeds that produced
by star formation by about one to two orders of magnitude. 

The AGN could therefore unbind the gas at least in some of the
sample. This is also shown in the middle left panel of
Fig.~\ref{fig:sfratio}, where we show the kinetic energy of the gas in
each AGN host galaxy as a function of the bolometric AGN
luminosity. The red solid line shows where the ratio of both is
 unity, and the dark blue dashed lines indicate conversion efficiencies
 between 0.1 and 10\% between bolometric luminosity and gas kinetic
 energy. The dark blue line shows the 5\% efficiency suggested
 by the black-hole bulge mass relationship. If this efficiency is
 approximately correct, then the AGN should provide sufficient energy
 to unbind the gas in about two thirds of our sample.

It is, however, still a matter of active debate amongst observers as
well as modelers whether feedback efficencies as high as few per
cent are not unrealistically high. Observations of molecular gas in
very dense, vigorously star-forming circumnuclear environments of
nearby AGN/starburst composites suggest that similarly high
conversions are necessary, if the AGN is to dominate the observed gas
kinematics. In turn, observations of UV absorption lines in FeLoBALs
only find efficiencies between 0.1 and 1\% \citep[][]{moe09, dunn10,
  bautista10}. FeLoBALs are considered to trace winds with
particularly high dust column densities, where the energy deposition
efficiencies should thus be high over comparably large radii. In
Fig.~\ref{fig:sfratio} this range corresponds to the light blue dashed
region. We label all galaxies that fall into or below this region
  in this diagram, and will return to a more detailed discussion of
  their properties below (\S\ref{sec:comparison}).

In addition to the energy transfer, the quasar radiation also injects
momentum into the gas. Each time a photon is scattered on a gas
or dust particle, a momentum of $h\nu/c$ is transferred, leading to a
total momentum transfer of $\xi L_{bol}/c$, where the 'momentum boost'
$\xi$ corresponds to the number of scatterings suffered by a photon
before it escapes, depending on the wavelength of the photon and the
optical depth along the path.

In the middle right panel of Fig.~\ref{fig:sfratio} we show the
momentum of the gas as a function of the momentum from the quasar
radiation, for different assumptions of $\xi$ between $\xi=1$ and
$\xi=20$.  Most SPH models find that momentum transfer of up to
$20\ L_{bol}/c$ is necessary to explain efficient AGN feedback as
implied by the black-hole bulge scaling relations, but make simplified
assumptions on, e.g., the radiative transfer, or the energy and
momentum transfer from the hot AGN wind onto the ambient gas. The
recent models of \citet{zubovas12} and \citet{faucher12}, which
combine the effects of the acceleration of gas in an expanding
blastwave and radiation pressure, suggest a momentum boost of about
$\xi=10$, highlighted as blue solid line in Fig.~\ref{fig:sfratio}.

Analytical work and hydrodynamic models taking radiative transfer
explicitly into account find however, that most AGN host galaxies,
except perhaps in the densest star-forming regions in ULIRGs, globally
may only reach a much lower total momentum transfer of about
$L_{bol}/c$ \citep[red line in Fig.~\ref{fig:sfratio},
  e.g.,][]{murray05, novak12, krumholz13}. In an adaptive-mesh
simulation with a simplified, but nonetheless explicit, radiation
transfer method, \citet{bieri16} recently found a rapid decrease from
about 30 to less than $1\times L_{bol}/c$ within the first $10^7$ yrs
after the ignition of the radiative AGN. The reason is that the
radiation escapes relatively easily after the AGN has blown
low-density channels into the dense gas clouds by which it is
initially enshrouded. This is also more in line with the observations
of FeLoBALs. Likewise, \citet{veilleux13} suggested that molecular
outflows from the highest-density inner regions of nearby ULIRGs
harbouring powerful, dust-enshrouded AGN, which are likely driven
mainly by radiation pressure, may subside once the AGN has cleared a
path through the ambient dust and gas clouds. Nonetheless, radiation
pressure and disk winds could perhaps play a role during early
feedback phases in our galaxies, before the winds have broken out from
the direct surroundings of the AGN in the central regions of the host
galaxy. While a momentum boost of $\xi=10-20$ would appear
  sufficient to power the observed outflows in about half of our
  sample, with $\xi=1$ the obscured quasars in our galaxies would be
  too weak for all but three sources with measurements of L$_{bol}$ (and
  one with an upper limit) to explain the observed gas kinematics. We
  have labeled these galaxies individually in Fig. ~\ref{fig:sfratio}, and
  will discuss their specific properties in more detail in
  \S\ref{sec:comparison} after examining the role of the radio jets.

The same is found when comparing the AGN bolometric luminosities in
our galaxies with those required to balance the gravitational
potential of the host galaxy through radiation pressure (but not
necessarily, to remove the gas through an outflow). \citet{murray05}
introduced such a quantity for a model of the combined feedback from
the radiation produced by AGN and star formation. They find that in
the most fortuitous case, when the surroundings of the AGN are
optically thick, the AGN must emit a critical luminosity $L_c = 4 f_g
c \sigma^4 / G$, which depends on the gas fraction, $f_g$, and stellar
velocity dispersion, $\sigma$. $G$ is the gravitational constant.

\begin{figure}
\begin{center}
\includegraphics[width=0.48\textwidth]{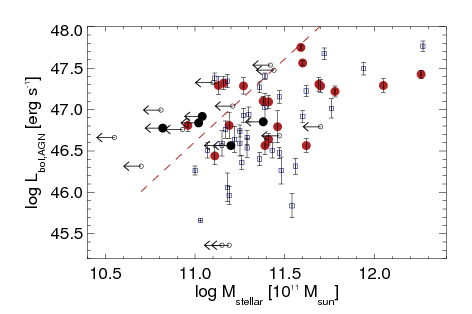}
\caption{Critical AGN luminosity for
  radiatively-driven quasar winds in the model of \citet{murray05} as
  a function of stellar mass. Large filled red dots and filled black
  dots show galaxies from our sample with measurements and upper limit
  of stellar mass, respectively. Small empty blue boxes and small
  empty black dots show the HeRGE galaxies without SINFONI
  observations and with measurements and upper limits on stellar mass,
  respectively. Quasar luminosities were taken from \citet{drouart14},
  and correspond to their AGN component of the FIR luminosity, which
  was scaled by a factor~6 to approximate the bolometric QSO
  luminosity. The dashed red line shows the critical luminosity at
  which the momentum carried by the quasar radiation may balance
  gravity \citep[see][for details]{murray05} \label{fig:qsolcrit}}
\end{center}
\end{figure}

Using the stellar mass estimates of \citet{seymour07} and
\citet{debreuck10} of typically few $10^{11}$ M$_{\odot}$, and the
molecular, atomic, and warm ionized gas mass estimates of, e.g.,
\citet{emonts14}, \citet{gullberg16}, and \citet{nesvadba16}, we find
that the typical gas fractions in HzRGs are about 10\%. Adopting the
virial theorem to estimate $\sigma$ from the same stellar masses by
setting $M=c\sigma^2\ r_e/G$, we find $\sigma=300-350$ km s$^{-1}$ for
galaxies with $M_{stellar}=2-4\times 10^{11}$ M$_{\odot}$ \citep[see
  also][]{nesvadba11b}. $c$ is a constant that depends on the radial
mass profile. We use $c=5$ \citep[][]{bender92}, and adopt $r_e=2-3$
kpc for the effective radius of the galaxy. In Fig.~\ref{fig:qsolcrit}
we show where the bolometric AGN luminosity of our galaxies falls
relative to the stellar mass and the critical luminosity implied by
\citet[][red dashed line]{murray05}. All but seven~sources have AGN
that would not be luminous enough to exceed the critical luminosity to
unbind the gas.

\section{Radio jets}
\label{ssec:jets}

Centimeter radio jets are powered by the synchrotron emission from
relativistic particles, with an emissivity that depends on the
surrounding gas density and magnetic field strength, and is difficult
to quantify accurately from first principles. Therefore, a number of
empirical measures have been developed to estimate the kinetic power
of a radio jet from the observed monochromatic radio power at a given
frequency. To estimate the global impact of the radio jets on the gas,
we rely upon the relationship of \citet{cavagnolo10}, who measured the
mechanical work done by radio jets in massive galaxy clusters to
inflate cavities in the intracluster medium, thereby providing
empirical estimates of the work done by the jet against the
surrounding gas.

The radio power of the entire SINFONI sample, including the 24
galaxies discussed here, has already been derived by
\citetalias{nesvadba16} by interpolating the observed multi-frequency
radio fluxes in the NASA/IPAC Extragalactic Database NED. Results for
individual targets are given in their Table~3. We summarize here only
the global results for our present targets, in form of the green
histograms in Fig.~\ref{fig:histoluminosities}. The left panel shows
the distribution of the radio power at rest-frame 1.4~GHz. To ease the
comparison with the far-infrared luminosity from star formation and
AGN, we also measure the radio luminosities in solar units, finding
that our sources fall into a range of $\log{P_{1.4}}=10^{10}-10^{12}$
L$_{\odot}$ (which corresponds to $10^{27-29}$ W Hz$^{-1}$ at 1.4~GHz
in the rest-frame). Our sources include thus some of the most powerful
radio sources known at their redshift. \citetalias{nesvadba16} also
demonstrate that these estimates are in good agreement with the
empirical calibration of, e.g., \citet[][]{willott00}, and also
detailed studies of the X-ray and radio spectral properties in two
nearby isolated radio galaxies \citep[][]{harwood16}. They are
conservative estimates, in that they predict a jet kinetic power that
is about $0.2-0.3$~dex lower than estimates by \citet{turner15}. The
latter take into account that the radio emission should become fainter
as the radio lobes expand \citep[see also][]{kaiser97}.

The right panel of Fig.~\ref{fig:histoluminosities} shows the kinetic
power of the radio jets obtained with the \citet{cavagnolo10}
approach. The values are about an order of magnitude greater than
those of the bolometric quasar power, and about three orders of
magnitude greater than the kinetic power of the star formation.

In the {\bf lower} left panel of Fig.~\ref{fig:sfratio} we compare the
kinetic energy of the gas with that provided by the radio jet. The
solid red and blue dashed lines in the right panel show energy
conversion efficiencies between 0.1 and 100\%. The jets provide
sufficient kinetic energy to explain the gas kinematics in all
sources, with a wide range of efficiencies. Since this energy comes
from an extended radio jet, it is being deposited locally into the
interstellar gas, which should make the energy transfer particularly
efficient \citep[e.g.,][]{mukherjee16}.

To estimate the momentum carried by the radio jet, we use the
calibration of \citet{cavagnolo10} already discussed above, and use
$p_{jet}= E_{jet} / v$, with $v =\beta c$.
Observations of compact steep spectrum radio galaxies in the nearby
Universe, which are good candidates of being in an early phase of jet
activity, where the jet interacts strongly with the dense
circumnuclear gas, suggest $\beta=0.01-0.1$
\citep[e.g.,][]{readhead96, owsianik98, harwood13}. Most suitable for
our purposes is to provide firm lower limits to the momentum injection
from the jets, and we therefore adopt $\beta=0.1$. Keeping in mind
that the observed velocities of entrained clouds are lower than the 
expansion velocity of the hot cocoon gas by at least factors of a few
\citep[][]{cooper08,gaibler09}, and that we observe velocities of up
to $\sim 2500$ km s$^{-1}$ for the warm ionized clouds in our SINFONI
sample, jet expansion velocities much smaller than $0.01c$ do not seem
very likely, unless the medium is very dense or clumpy, so that the
jet has difficulties to escape. Hydrodynamic models, however, show,
that during such phases jets are particularly efficient in imparting
their momentum into the gas. As the jets push through low-density
channels in such galaxies, their ram pressure produces mechanical
advantages of factors 10-100, so that the velocities of the entrained
gas may reach velocities of few 10$^{2-3}$ km s$^{-1}$, as observed
\citepalias{nesvadba16}, although the jet expansion velocity itself
may actually be much lower \citep[][]{wagner11,wagner12}. As a result,
the galaxies would fall into a very similar range in
Fig.~\ref{fig:sfratio}, although the physics behind the gas
acceleration would be more complicated. 

The momentum of the gas as a function of the momentum provided by
  the radio jet in our sources is shown in the bottom right panel of
  Fig.~\ref{fig:sfratio}. The black dashed region in
  Fig.~\ref{fig:sfratio} shows the range in the diagram where a jet
  expansion velocity less than $0.01 c$ would be required, which we
  consider unlikely. All galaxies fall below this range, i.e., within
  a region of the diagram where the momentum of the gas can be
  plausibly powered by the radio source. We note that galaxies with
  greater jet expansion velocity transfer less of their momentum into
  the gas, so that it is not unphyisical to find galaxies which fall
  formally in the range where $v_{jet}>c$. In this case, additional
  effects, perhaps due to geometry, or clumpiness of the gas, could
  make the momentum transfer less efficient than implied by our
  simplified assumptions.

\section{Ambiguous cases}
\label{sec:comparison}

Our sample includes six targets, for which the simple energy and
momentum considerations presented above do not allow to distinguish 
between AGN bolometric and radio power as the main driver of the gas
kinematics. Both radio source and quasar radiation seem powerful
enough to explain their observed gas kinematics (either kinetic
energy, momentum, or both), even when adopting the relatively low
energy and momentum transfer efficiencies suggested by FeLoBAL
observations and the most detailed hydrodynamic models
\citep[][]{moe09, dunn10, bautista10, krumholz12, bieri16}.

These six targets are amongst those for which we measured the lowest
kinetic energy and momentum in our SINFONI data sets. Three of these
targets, MRC~2224$-$273, MRC~0114$-$211, and TN~J2007$-$1316, are not
well spatially resolved, so that beam smearing might lower their
velocity measurements and thus their kinetic energy and momentum
estimates. This is in particular the case for TN~J2007$-$1316, which
has one of the lowest velocity gradients measured in our entire
sample, $\Delta v= 100$ km s$^{-1}$
\citepalias[][]{nesvadba16}. PKS~0529$-$549 and 3C~257, in contrast,
have well resolved emission-line regions, and radio jets that are
still confined within the extended emission line gas. Both galaxies
show kinematic disturbances associated with the extended radio
emission, which suggests that most of the energy injection comes from
the radio source.

\begin{figure*}
\begin{center}
\includegraphics[width=0.49\textwidth]{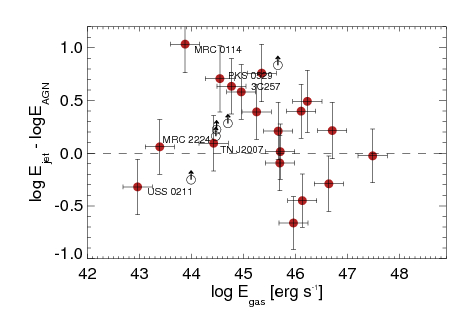}
\includegraphics[width=0.49\textwidth]{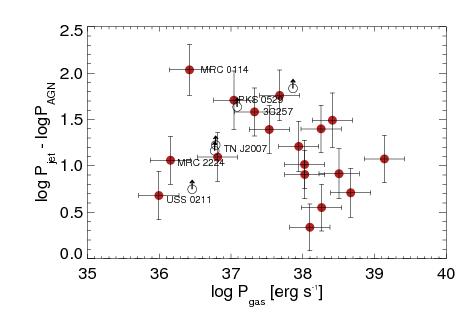}
\caption{\label{fig:energyratios}
Ratio of the kinetic energy {\it (left)} and momentum {\it (right)}
 transfered into the gas through AGN radiation and radio jet as a
 function of those observed in the gas. The dashed line shows where
 both are equal. We label individual targets, for which a simple
 comparison of the energy and momentum input rates do not allow to
 either favor radiation or jets as main injection mechanism.}
\end{center}
\end{figure*}

The sixth source, USS~0211$-$122 has very low line widths in the
extended gas, and pronounced blue wings in the
[OIII]$\lambda\lambda$4959,5007 line profiles
\citepalias[][]{nesvadba16}. This is unusal compared to the remaining
sample of HzRGs with SINFONI data from \citetalias{nesvadba16}, where
line profiles are overall broad, reflecting the large line widths and
strong velocity gradients over most of the bright emission-line gas,
but do not show pronounced secondary components, as commonly found in
quasars. \citet{collet15b} propose that this could be an intrinsic
difference between radio galaxies and quasars, perhaps indicating
different radial sizes of the outflows. If this is correct, then
USS~0211$-$122 could well be a source where the energy injection from
the bolometric quasar radiation exceeds over that from the radio
jet. This could also explain why surveys of large samples of typically
unresolved quasar spectra find that broad, blueshifted wings appear
more associated with radio-quiet than radio-loud AGN
\citep[][]{heckman84, greene05, woo16}, whereas detailed observations
of individual galaxies with radio-loud AGN find perturbed kinematic
features preferentially along the radio jet axis
\citep[e.g.,][]{husemann13}.

In Fig.~\ref{fig:energyratios} we show the ratio between the energy
and momentum injection from radiation and jets as a function of gas
kinetic energy. The figure shows that the expected energy and momentum
injection related to the jet is typically higher by at least factors
of a few than those from the AGN radiation. MRC~0114$-$211,
PKS~0529$-$549 and 3C~257, are in fact amongst the galaxies, where the
power of the radio jet is particularly strong compared to the
bolometric power. It would not be obvious why the gas kinematics in
these sources should be dominated by the radiation rather than the
jet, if the opposite is the case in most of the other sources, as
suggesed by Fig.~\ref{fig:sfratio}. In USS~0211$-$122, however, the
source with particularly high rest-frame UV polarization
\citep[][]{vernet01} already discussed in \S\ref{ssec:starformation},
the bolometric power is greater than the radio power, and the
mechanism powering the gas kinematics remains therefore more
ambiguous.
 
\section{Discussion and conclusions}
\label{sec:conclusion}
 
We used observations of the kinematics of warm ionized gas, and the
far-infrared and centimeter radio spectral energy distribution of 24
powerful radio galaxies at z$\sim$2, to compare the rates of kinetic
energy released by star formation, quasar radiation, and radio jets,
with the energy input rates required to power the observed gas
kinematics. All galaxies have far-infrared spectral energy
distributions measured with Herschel/PACS and SPIRE photometry,
which \citet{drouart14} decomposed into an AGN and star-formation
component (decompositions for individual targets are shown in their
Fig.~D.1). They have also been observed with SINFONI to infer the
kinematics of the warm ionized gas \citep[][]{nesvadba16}. We use
standard techniques from the literature to estimate the kinetic energy
and momentum released by star formation, radiation from the buried
AGN, and radio jet, and to compare with the kinetic energy and
momentum found in the gas. We also take into account that the
efficiency of the energy and momentum transfer from each source to the
gas is unlikely to be unity, and compare with several assumptions
given in the literature. Our understanding of AGN feedback is still
evolving very rapidly, and our goal was to capture the main streams of
the current discussion in our analysis, while also providing
observational constraints that are generic enough to be useful for
comparison with future theoretical work.

We find that the kinetic energy produced by starburst-driven winds
falls short of what is required to drive the observed gas kinematics,
in agreement with previous results for individual sources
\citep[][]{nesvadba06a, seymour12}. For radiative AGN feedback
(``quasar feedback'') the potential of producing the observed gas
kinematics is more ambiguous, and depends on the assumed efficiency
with which the gas is being accelerated. For the most optimistic
assumptions, AGN radiation could be behind the gas kinematics in about
two-thirds of our sources. For models including more explicit
descriptions of the radiative transfer, lower efficiencies are
found. In this case, the quasar emission would generally not be
sufficient to power the kinematics of the warm ionized gas. In
six galaxies with low gas kinetic energy and momentum, AGN radiation
could suffice to power the kinematics even with these relatively low
injection rates, however, the specific properties of these sources
suggest nonetheless that the radio jet is probably more
effective. One notable exception is USS~0211$-$122, which has
unusual [OIII] line profiles more reminiscent of quasars than radio
galaxies, and which also has a low expected rate of jet to AGN
energy and momentum.

Radiation pressure seems also insufficient to unbind the gas from
galaxies with stellar masses as high as in our targets. The radio jets
appear capable to providing sufficient energy and momentum to produce
the highly energized, extended emission-line regions we observe in
these galaxies.

The analysis presented here is complementary to that of
\citet{nesvadba16}, who analyzed the detailed properties of 49 HzRGs,
including the 24 we discuss here, which also have the Herschel
far-infrared photometry necessary for a detailed discussion of the
energy produced by AGN and star formation. They found several
signatures, e.g., a general co-alignment of radio jet axis and major
axis of the extended emission-line regions, and abrupt changes in
velocity or line widths associated with the jet direction or radio
features, which also suggest that the radio jet is the main source of
energy and momentum injection. 

We emphasize that our results do not imply that AGN radiation and star
formation have no effect at all on their surrounding gas.  With an
analysis of the global gas properties as done here, we cannot probe
further than identify the most powerful mechanism. Nonwithstanding,
our results do not require us to assume more complex scenarios, e.g.,
a conspiracy of gas acceleration through radiation and jets, as would
be the case if no single mechanism was able to provide sufficient
momentum and energy by itself. Detailed, high-resolution studies of
radio-loud quasars at low redshift currently paint a mixed picture
with some authors finding that radiation dominates
\citep[e.g.,][]{liu13}, and others who attribute a larger role {\bf to} the
radio jet \citep[e.g.,][]{husemann13}.

Our results for this particular moment in the evolution of our sources
are also independent from time scale or duty cycle considerations,
because we observe the impact of each source in-situ and at the same
moment acting on the same gas reservoir. With jet lifetimes of few
$10^{6-7}$ yrs \citep[e.g.,][]{blundell99}, and star-formation
timescales of at least few $10^8$ and perhaps even up to $10^9$ yrs
\citep{rocca13}, it is clear that we are witnessing these processes at
a peculiar moment in time, likely, when the active formation phase of
the host galaxy is nearly terminated \citep[e.g.,][]{nesvadba11a,
  drouart14}. While star formation and AGN radiation have probably
affected the host evolution prior to this phase, they have not removed
the gas we see in these targets now. Our results therefore suggest
that the radio jet dominates the final 'sweeping clean' of gas from
these very massive galaxies. This is also consistent with the result
presented in Fig.~\ref{fig:qsolcrit} that AGN with radiative power of
$10^{46-47}$ erg s$^{-1}$, in the range of the most luminous AGN at
these redshifts, do not seem able to overcome the gravitational
potential of galaxies with $M\ge 10^{11}$ M$_{\odot}$ through the
momentum carried by their bolometric radiation alone. In earlier
stages of assembly of our galaxies, and lower-mass galaxies generally,
the relative balance of star formation, AGN radiation, and radio jet,
might therefore be very different.

Making this comparison for the present set of galaxies was relatively
simple, because the sources are well extended, and the radio sources
are bright enough that their kinetic power dominates globally over
other mechanisms. However, as one moves down the radio luminosity function,
this becomes more difficult, as can be seen, e.g., from a very similar
analysis performed by \citet{nesvadba11b} and \citet{polletta11} for
two of the brightest obscured quasars from the Spitzer SWIRE survey
\citep[][]{polletta08}. The obscured quasars in these sources are
about as powerful as the ones we discuss here, but their radio sources
are about four orders of magnitude fainter than the brightest sources
we discuss here \citep[][]{sajina07}. Nonetheless, their gas
kinematics seems to be dominated by the radio jet, contrary to AGN
selected, e.g., in the optical or X-ray. Much deeper and
more detailed analyses with high-resolution radio data, will be
required to study galaxies where the energy output from star
formation, quasar, and jet are more balanced. However, this will be a
long-term effort, as currently, no general, reliable, uncontested
methods exist, e.g., to distinguish between star formation and AGN in
faint high-redshift radio sources. ALMA, the next generations of ELTs
and JWST in the optical and near-infrared, and radio facilities like
LOFAR and SKA, will undoubtedly play a large role in disentangling the
relative contribution of each process for the regulation of the gas
kinematics in the general population of high-redshift AGN host
galaxies.

\acknowledgements
We thank the anonymous referee for insightful comments that helped
improve the paper, and the staff at Paranal observatory and at ESA for
their assistance with obtaining the data on which this work is
based. NPHN also wishes to express her gratitude to M.~D.~Lehnert for
very interesting and inspiring comments on the draft of this paper,
and his continuous advice and guidance in the more than 10 years of
previous collaboration, that also form the basis of the present
analysis. She wishes to thank ESO for supporting this study through
DGDF grant 14/10, and for the hospitality of the staff in Garching
during the extended visit that ensued. PNB is grateful for support
from the UK STFC via grant ST/M001229/1.

The {\it Herschel\/} spacecraft was designed, built, tested, 
launched and operated under a contract to ESA managed by the {\it
  Herschel/Planck\/} Project team by an industrial consortium under
the overall responsibility of the prime contractor Thales Alenia Space
(Cannes), and including Astrium (Friedrichshafen) responsible for the
payload module and for system testing at spacecraft level, Thales
Alenia Space (Turin) responsible for the service module, and Astrium
(Toulouse) responsible for the telescope, with in excess of a hundred
subcontractors.

\bibliographystyle{aa}
\bibliography{hzrg}

\end{document}